\documentclass[conference]{IEEEtran}
\usepackage{cite}
\usepackage{enumitem}
\usepackage{amsmath,amssymb,amsfonts}
\usepackage{algorithmic}
\usepackage{graphicx}
\usepackage{textcomp}
\usepackage{xcolor}
\def\BibTeX{{\rm B\kern-.05em{\sc i\kern-.025em b}\kern-.08em
    T\kern-.1667em\lower.7ex\hbox{E}\kern-.125emX}}

\usepackage{fancyhdr}
\newcommand{\m}[1]{\mathcal{#1}}

\begin{document}

\title{A Heterogeneous Multimodal Graph Learning Framework for Recognizing User Emotions in Social Networks\\
}

\author{\IEEEauthorblockN{Sree Bhattacharyya, Shuhua Yang, James Z. Wang} 
\IEEEauthorblockA{\textit{College of Information Sciences and Technology} \\
\textit{The Pennsylvania State University, University Park} \\
\textit{\{sfb6038, sky5341, jzw11\}@psu.edu}}
}

\maketitle
\thispagestyle{fancy}

\begin{abstract}
The rapid expansion of social media platforms has provided unprecedented access to massive amounts of multimodal user-generated content. Comprehending user emotions can provide valuable insights for improving communication and understanding of human behaviors. Despite significant advancements in Affective Computing, the diverse factors influencing user emotions in \textit{social networks} remain relatively understudied. Moreover, there is a notable lack of deep learning-based methods for predicting user emotions in social networks, which could be addressed by leveraging the extensive multimodal data available. This work presents a novel formulation of personalized emotion prediction in social networks based on heterogeneous graph learning. Building upon this formulation, we design \textit{HMG-Emo}, a \underline{H}eterogeneous \underline{M}ultimodal \underline{G}raph Learning {F}ramework that utilizes deep learning-based features for user \underline{emo}tion recognition. Additionally, we include a dynamic context fusion module in HMG-Emo that is capable of adaptively integrating the different modalities in social media data. Through extensive experiments, we demonstrate the effectiveness of HMG-Emo and verify the superiority of adopting a graph neural network-based approach, which outperforms existing baselines that use rich hand-crafted features. To the best of our knowledge, HMG-Emo is the first multimodal and deep-learning-based approach to predict personalized emotions within online social networks. Our work highlights the significance of exploiting advanced deep learning techniques for less-explored problems in Affective Computing.
\end{abstract}


\section{Introduction}

\begin{figure}[tbp]
    \centerline{\includegraphics[scale=0.48]{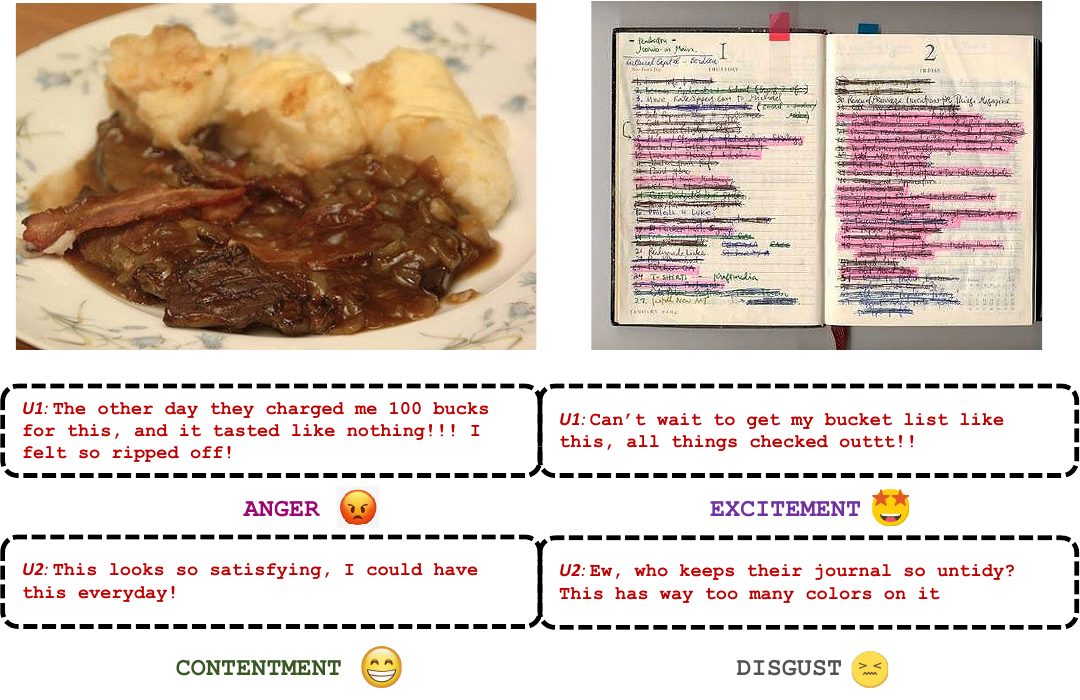}}
    \caption{Example from the IESN~\cite{zhao2016predicting} dataset. The emotion labels are dependent on both the image contents and the user comments, showing the complex nature of emotion interpretation.}
    \label{fig:1}
\end{figure}

Advancements in deep learning and artificial intelligence over the last two decades have sparked significant research interests in methods for automatic emotion recognition that utilize multimodal data. Existing research in emotion recognition can be broadly categorized into evoked and expressed emotion recognition~\cite{wang2023unlocking}. 
Evoked emotion refers to the emotions elicited in individuals when interacting with certain external stimuli, regardless of whether those emotions are explicitly expressed. 
Expressed emotion recognition, on the other hand, regards individuals as the source of specific emotional actions, conveying their feelings through facial expressions, verbal expressions, or body movements. 
When studying evoked emotion, researchers often focus solely on the external stimuli that prompt the corresponding emotional change. However, in real-life scenarios, evoked emotion can be a consequence of multiple factors that have a simultaneous impact. 
Understanding emotions within such complex contexts requires the integration of diverse information from multiple modalities or contexts, leading to the development of context-aware emotion recognition (CAER) tasks.

While most CAER methods for evoked emotion recognition focus on contextual information from the external stimuli, such as historical context in conversations~\cite{ghosal2019dialoguegcn}, scene and object information in images~\cite{nagappan2023context, peng2021affect}, or a fusion of multiple modalities from video data~\cite{mittal2020emoticon}, there remains a gap in considering contexts from the individuals whose emotions are evoked. 
This information, however, is particularly critical in social networks, where user contexts play a pivotal role. 
In social media platforms, user-specific context can be extremely important, as individual users coming from various backgrounds constantly interact with visual content on the platform and actively get influenced by their connections with other users~\cite{yang2014how}. The challenge in emotion recognition is further amplified by the possibility of varied interpretations of the same media by different individuals, leading to personalized evoked emotions. 
For instance, consider the scenario depicted in Fig.~\ref{fig:1}. 
We can observe that the emotions evoked by the same image can vary among different users. It is only when contextual cues, such as user comments alongside the images, are taken into account that personalized emotion assessments become meaningful. Additionally, leveraging user-related information, such as social ties, can provide valuable context, as community dynamics often significantly influence individual emotions~\cite{jin2017emotions}.

Investigating personalized emotions within social networks can potentially benefit numerous downstream tasks. To start with, fine-grained knowledge of user behaviors in social media has multifarious applications such as content recommendations, polarity detection, and content moderation~\cite{ravi2015survey}. Since emotions are often associated with opinion mining~\cite{zhao2016book, kumar2015emotion}, examining emotions within social media platforms can also be used to gauge user sentiment on global matters. Additionally, the evolving emotional states of users over time are important for identifying social media-related mental health issues, especially in unusual circumstances like the pandemic, which saw a sharp spike in digital engagement~\cite{yang2020social}.

Several researchers utilized social media data to investigate affective information drawing on various modalities of multimedia content~\cite{yang2017learning}. However, few approaches address the problem from a user-centric perspective, predicting personalized emotions evoked in users~\cite{zhao2016predicting,yang2014how}, by including information directly associated with users as an influencing factor. Existing user-centric methods rely only on low-level hand-crafted features from different modalities and probabilistic or deterministic graph modeling for personalized emotion prediction. The power of graph-based deep learning has not been exploited adequately to address personalized emotion recognition for users in social networks.
Therefore, in this work, we present the first deep graph learning-based framework for personalized user emotion recognition in social networks. We create a framework that allows features learned from different modalities to be combined adaptively in the form of a user-media graph, wherein the features are further refined through graph learning. We utilize the only public dataset for personalized emotion prediction~\cite{zhao2016predicting} to test the effectiveness of our framework, as it includes information about users and the images they upload and interact with. We also include a comparison of our method with the existing baseline~\cite{zhao2016predicting} that employs multi-task hypergraph learning.
Our main {\bf contributions} can be summarized as follows: 
\begin{itemize}[left=0pt]
    \item We introduce a novel formulation for the problem as an edge classification task in a heterogeneous user-image graph, which enables the intuitive and easy use of advanced graph learning methods to approach personalized emotion recognition in social networks.
    \item We create \textbf{HMG-Emo}, a \underline{H}eterogeneous, \underline{M}ultimodal, \underline{G}raph Learning framework that utilizes an adapted Graph Attention Network~\cite{velivckovic2018graph} for \underline{emo}tion recognition. It accumulates information from multiple modalities simultaneously and includes a plug-and-play dynamic context combination module to attend adaptively to different modalities during the emotion classification task.
    \item With thorough experiments on the well-established public dataset in this domain~\cite{zhao2016predicting}, we verify that our framework outperforms the existing baselines for emotion classification. Furthermore, our method does not require high-level, hand-crafted features. We also demonstrate the effectiveness of using multiple contextual sources of information and the adaptive combination module. 
    \item Through extensive ablation experiments, we further validate the robustness of different components in the proposed framework. We also generatively augment part of the benchmark dataset, to examine the importance of using multiple factors in emotion prediction. 
\end{itemize}

\section{Related Work}
\label{sec2}
In this section, we highlight recent related work on context-aware and multimodal emotion recognition and introduce studies of emotion in the context of social networks. 
\begin{figure*}[tbp]
    \centerline{\includegraphics[scale=0.54]{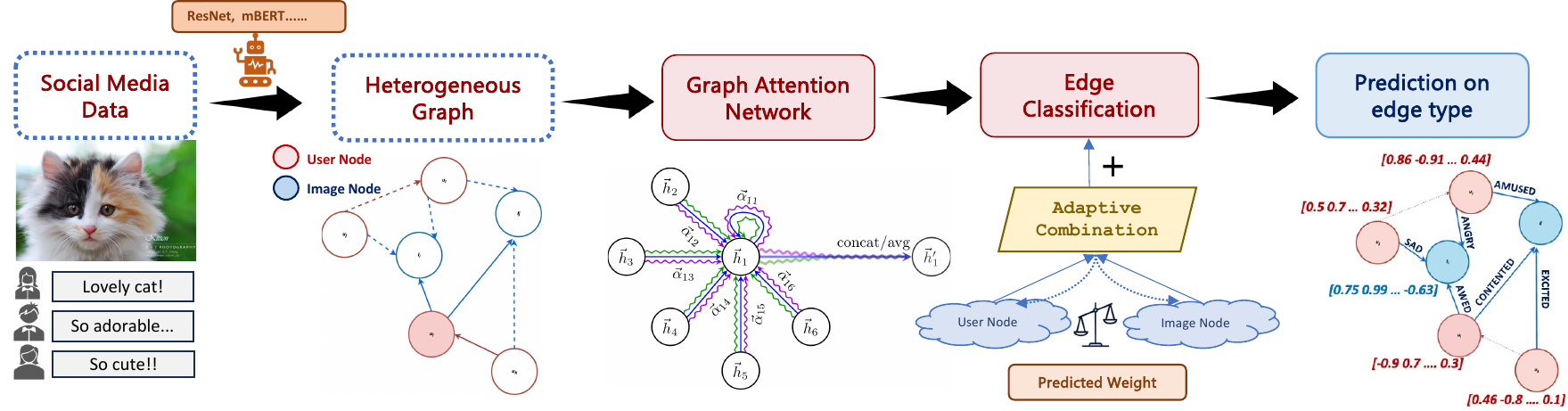}}
    \caption{An overview of the complete methodology pipeline.}
    \label{fig:2}
\end{figure*}

\subsection{Emotion Recognition: Context-Aware and Multimodal} 
CAER has seen the use of both unimodal and multimodal data for automatically recognizing emotions with a situated perspective. Among published work that focuses on using the \textit{visual modality} singularly, recent approaches adopt the fusion of contextual information in various ways, such as through expanding the original emotion label space into a combined emotion-context matrix~\cite{balouchian2018context}, using information from the full image along with other specific components like the human body or a segmentation map of an image~\cite{peng2021affect}. More recent approaches use background knowledge in the form of spatial information~\cite{hoang2021context}, object and scene information~\cite{nagappan2023context}, and human body features~\cite{khargonkar2023selinet}. 

In the \textit{modality of text}, applications such as emotion recognition in conversations (ERC) have seen the use of historical information from the conversation as context, using self and inter-speaker dependency to guide the task of identifying emotions~\cite{ghosal2019dialoguegcn}. Recently, several unique approaches have exploited context dependency in conversations. SACL~\cite{hu2023supervised} uses contextual negative samples for improved robustness in ERC, and another approach learns separate features based on whether a particular training instance should be considered context-dependent or independent to mitigate any impact from noisy contextual information~\cite{tu2023context}. Recent work has also included meta-information such as discourse structure in conversations~\cite{zhang2023dualgats} for emotion recognition.

For \textit{video data}, a wide variety of contextual information has been used. This includes using features such as the appearance and motion of subjects in videos~\cite{yin2016video}, face and gait information, along with background scene information and depth maps for video frames~\cite{mittal2020emoticon}. Other approaches have used dense optical flow~\cite{pikoulis2021leveraging}, body language and movements~\cite{filntisis2020emotion}~\cite{luo2020arbee}~\cite{dael2011emotion}~\cite{gunes2006bimodal}, or facial expressions~\cite{zheng2023facial}. The development of video-based context-aware datasets~\cite{kosti2019context} has also propelled the growth of such methods. However, recent years have seen a marked shift to using all modalities in conjunction, to create an emotion recognition pipeline that resembles the human process of understanding emotions even more closely.

When emotion recognition uses \textit{multiple modalities}, each modality could be considered to serve as complementary contextual information for the other modalities. Relatively earlier approaches in using multimodal features simultaneously include the introduction of a Tensor Fusion Network to learn intra-modality and inter-modality dynamics for inputs from three different modalities~\cite{zadeh2017tensor}, and models with gating units storing speaker-specific, and multimodal, information in ERC~\cite{hazarika2018conversational}. More recent approaches have focused on the complexities of the interaction of multiple modalities. This includes the use of cross-modal context fusion networks~\cite{zhang2023cross}, pretrained multimodal models for feature extraction~\cite{bose2023contextually}, contrastive learning on multimodal features~\cite{hu2022unimse}, and cross-modal attention-based methods~\cite{shi2023multiemo}.

\subsection{Emotion Recognition in Social Networks}
With social media being ubiquitous today, emotion recognition on social media content has received widespread attention in the field of affective computing.
There have been popular datasets for emotion recognition that utilize social media platforms as data sources, including Twitter~\cite{you2015robust, borth2013large, yang2017learning}, Instagram and Flickr~\cite{katsurai2016image}. However, most of these datasets do not consider user context, such as user preferences and connections.
Some approaches also attempt to examine emotions by combining multiple modalities such as text, images, and videos from social media~\cite{illendula2019multimodal, xu2018het}. Two earlier approaches attempt to model user emotions holistically from a user-centered view, considering information including the user contacts, groups, and sequential order of users viewing different images in Flickr, along with studying the multimedia content~\cite{zhao2016predicting, yang2014how}. One work uses hypergraph-based learning, devising a learning algorithm that explores the correlation between the different user relations and images~\cite{zhao2016predicting}, while the other work uses statistical modeling to study the effect of social relations on user affect. Essentially, the use of deep learning for user emotion recognition remains under-explored. Further, utilizing deep graph learning for multimodal and personalized emotion recognition in social networks remains an open challenge.

\section{Methodology}
\label{sec3}

In this section, we give a detailed description of the proposed formulation of the personalized user emotion prediction problem and our framework. Fig.~\ref{fig:2} provides a visual reference for the complete steps in the framework. 

\subsection{Problem Formulation}
\label{sec3-a}

Our goal is to develop a framework to predict personalized emotional responses from users during their interaction with visual media in a social network. We aim to include user contexts alongside features of the visual stimuli they interact with. This requires a robust framework that can combine multiple factors seamlessly to make reliable predictions for evoked personalized emotions in users. 

\begin{figure*}[ht!]
    \centerline{\includegraphics[width=15.6cm]{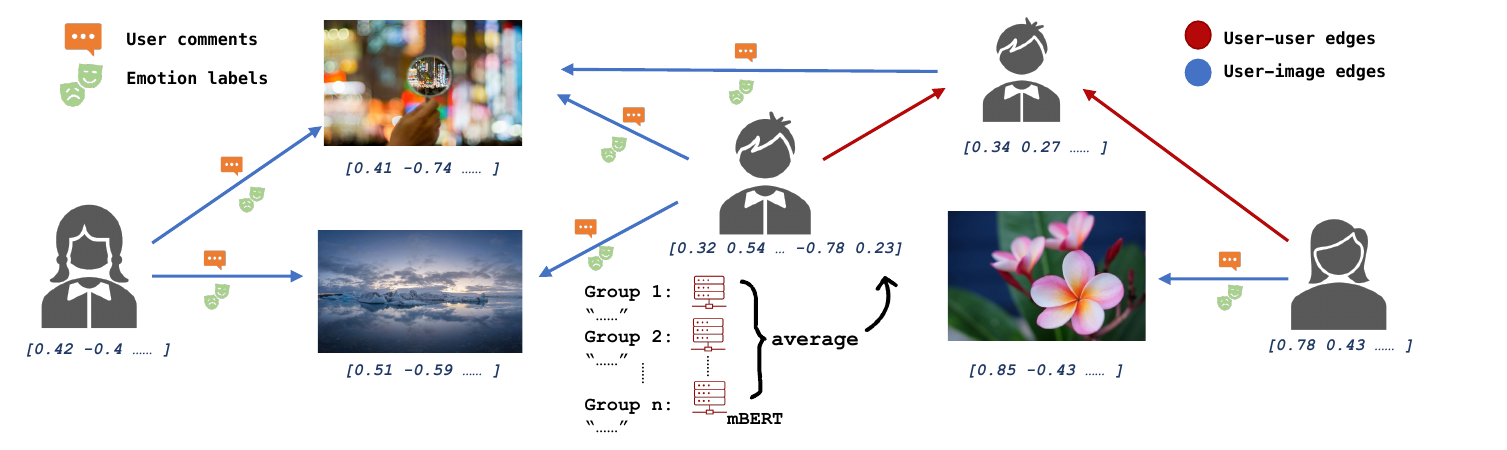}}
    \caption{The structure of the created heterogeneous graph. The feature generation process for a particular user is depicted in one of the user nodes.}
    \label{fig:3}
\end{figure*}


To formulate this problem, we consider two types of user interactions in a social network: interactions with other users, and interactions with the visual contents, namely, images. 
Therefore, we utilize two basic entities, the users and the images, to create a heterogeneous interaction graph that can model both interactions simultaneously. 
We define the heterogeneous graph \(\m{G} = (\m{V, E})\), where \(\m{V}\) represents the set of nodes containing two subsets: \(\m{V}_\text{user}\) and \(\m{V}_\text{image}\). 
The set of edges \(\m{E}\) encompasses two types of edges, denoted as edge type \textit{views} and edge type \textit{connects}. Here, \(\m{E}_\text{views} \subseteq \m{V}_\text{user} \times \m{V}_\text{image}\) represents connections between users and the images they view on the platform, while \(\m{E}_\text{connects} \subseteq \m{V}_\text{user} \times \m{V}_\text{user}\) corresponds to edges connecting users who are contacts or connections of each other. 

In our graph representation, the images are considered as independent nodes rather than features of the user nodes. 
We also embed additional information as node and edge features in \(\m{G}\). 
For example, the extracted image features can be included as the node features for all nodes of type \(\m{V}_\text{image}\). 
Likewise, the user comment \(i\) when viewing image \(j\) can be considered as a feature of the edge \(e_{i,j}\), where \(e_{i,j} \in \m{E}_\text{views}\). 
For user \(i\) and image \(j\), the evoked emotion ground-truth labels are considered as the supervision label for edge \(e_{i,j}\), indicating that the evoked emotion comes from viewing image \(j\). 
Thus, the task of predicting evoked emotions in users becomes a supervised edge classification task on a single type of edge in the graph (\(\m{E}_\text{views}\)), which is done by aggregating information from various sources using the graph structure, node, and edge features. 
This formulation presents two advantages. Firstly, the users and images exist as independent nodes, allowing for the creation of user features pertaining directly to each user, as described in the next section. The separation of the entities to constitute two different types of interactions also helps in effectively increasing the number of instances to use for modeling emotions. Secondly, this formulation makes advanced graph learning methods an intuitive choice to solve the problem. 

\subsection{Unimodal Processing of Contexts}\label{sec3-b}

We first process each source of contextual information to be included in HMG-Emo as follows: 
\begin{itemize}[left=0pt]
    \item \textit{Visual Context}: This refers to the actual visual media, or images, being viewed by the users on the social media platform. We use high-level features obtained using a pretrained Resnet-50~\cite{he2016deep} model, moving beyond the low-level, hand-crafted features that have been employed for this purpose in the past. The features obtained are then treated as node attributes for all nodes of the type \textit{image}.
    \item \textit{User Context}: The user context aims to capture information directly related to the user. This is subdivided into two parts: 
    \begin{itemize}
        \item \textit{User-Image Interaction Context}: We consider the comments left by users on images as the contextual information of a single user-image interaction. We include the comment information as edge features for edges connecting users and images. 
        \item \textit{General User Context}: Besides incorporating the information of user connections implicitly through the heterogeneous graph structure, we also include information about different special interest groups joined by users on a social media platform. Such groups on photo-sharing websites are usually joined by like-minded users interested in sharing images aligned to a particular theme. We use the textual descriptions of each such special interest group and obtain their features. Then, for every user, we aggregate the features obtained for all such groups the user belongs to, to be used as a node feature for user nodes in the graph. Formally, consider \(\m{S}\) to be the set of all possible special interest groups present in the social media platform. Now, for user \(i\), say \(u_{i}\) is a part of all groups in \(\m{S}_{i}\), where \(\m{S}_{i} \subseteq \m{S}\). Along with that, we have a text description \(t\), corresponding to each \(s \in \m{S}\). Thus, for user \(u_{i}\), we calculate the group context feature by the following formulation: 
        \begin{equation}
            g_{i} = \frac{1}{|\m{S}_{i}|}\sum_{s \in \m{S}_{i}}f(s)\;, \\
        \end{equation}
        where \(f\) is the feature extractor for each textual description \(t\) for all \(s \in \m{S}\).
    \end{itemize}
    For both the \textit{interaction-level context}, and \textit{general user context}, we thus need to use a text-based feature extractor. We use zero-shot features learned from multilingual-BERT~\cite{devlin2019bert}, to handle text in multiple languages. The features learned using comments are used as \underline{edge} features for user-image edges and the features learned from group descriptions are used as \underline{node} features for all user nodes. Wherever the comment texts are missing, we initialize the edge attribute with a uniform feature tensor.
\end{itemize}
The reason we rely on popular pre-trained models for zero-shot feature extraction is to ensure that the unimodal feature learning process can be as straightforward as possible, and not require any human supervision. As opposed to previously used methods, learning features from models in a zero-shot manner saves the requirement of domain knowledge for designing hand-crafted low-level features or spending time and resources on retraining models to obtain features, while also providing significant performance improvements in the downstream task of classifying evoked emotion. Fig. \ref{fig:3} shows the graph structure created, along with the process for creating node features for user nodes.

\subsection{Graph Learning}
\label{sec3-c}

\subsubsection{Graph Attention Network}

Starting with the heterogeneous graph, we include additional information as low-dimensional features learned in a zero-shot manner to add as node and edge attributes. To have an algorithm that aggregates all of this information and predicts edge labels for a single type of edge, we use a graph attention network backbone~\cite{velivckovic2018graph}, adapted to suit the structure of our heterogeneous graph. Essentially, the network should assimilate information from the graph structure, node, and edge features, in multiple steps or hops, and update the representation learned for the nodes. Based on the low-dimensional node representations, a classification problem is formulated, that predicts the emotion label for an edge using the representations of its source and destination nodes. We ensure that the learning mechanism uses edge attributes only for user-image edges. Further, it is important to note that the model should aggregate information over all types of nodes and edges while using supervision only for user-image edges that have ground truth emotion labels. Mathematically, considering a spatial view of the heterogeneous graph, for every layer of graph attention, the following is computed: 
\begin{equation}
    {x}^{\prime}_i = \alpha_{i,i}\mathbf{\Theta}{x}_{i} +
    \sum_{j \in \m{N}(i)}
    \alpha_{i,j}\mathbf{\Theta}{x}_{j}\;, 
\end{equation}
where \({x}^{\prime}_i\) represents the updated representation for node \(x_i\) and \(\alpha_{i,i}\) and \(\alpha_{i,j}\) are the attention coefficients for self and neighboring nodes of the node \(x_i\)~\cite{velivckovic2018graph}. \(\mathbf{\Theta}\) represents the learnable weight matrix for the node interactions, and can be considered the general weight matrix that transforms nodes into high-level feature representations. Specifically, the attention coefficient $\alpha_{i,j}$ is calculated as
\begin{eqnarray}
\frac{
\exp\left(f\left(
\mathbf{a}^{\top}_{s} \mathbf{\Theta}\mathbf{x}_i
+ \mathbf{a}^{\top}_{t} \mathbf{\Theta}\mathbf{x}_j
+ \mathbf{a}^{\top}_{e} \mathbf{\Theta}_{e} \mathbf{e}_{i,j}
\right)\right)}
{\sum_{k \in \mathcal{N}(i) \cup \{ i \}}
\exp\left(f\left(
\mathbf{a}^{\top}_{s} \mathbf{\Theta}\mathbf{x}_i
+ \mathbf{a}^{\top}_{t} \mathbf{\Theta}\mathbf{x}_k
+ \mathbf{a}^{\top}_{e} \mathbf{\Theta}_{e} \mathbf{e}_{i,k}
\right)\right)}\;,  
\end{eqnarray}
where \(f(\cdot)\) is a non-linearity like LeakyRelu, \(\mathbf{a}\) is the weight matrix associated solely with the attention mechanism, with \(\mathbf{a_s}, \mathbf{a_t}, \mathbf{a_e}\) denoting the weights for attention to self, neighbor and the edge connecting them, respectively. \(\mathbf{\Theta}_{e}\) is the learnable weight matrix for transforming the edge attribute features, used only for user-image edges. In HMG-Emo, each Graph Attention layer is followed by Batch Normalization~\cite{ioffe2015batch}, and a non-linearity of ReLU~\cite{nair2010rectified}.

Once the individual node representations are learned, they are combined using an adaptive mechanism and then subjected to a classification module. Precisely, beyond the unsupervised graph learning stage, we have the node features \(x_i \in \m{R}^\m{D}\), where \(\m{D}\) is the final dimension for the node features. Then, the classification task becomes to find the mapping:
\begin{equation}
    h: g(x_i, x_j) \rightarrow \m{C}\;,
\end{equation}
where \(g(\cdot)\) represents a mechanism to combine the node features of the source and destination node, and \(\m{C}\) represents the class of emotion labels. Note that this is carried out only for edges of the \textit{views} type. Our classification module consists of a simple feedforward network, consisting of 3 hidden layers, with non-linear activation in the form of ReLU~\cite{nair2010rectified}, and Dropout~\cite{srivastava2014dropout} layers after each of them, except the last.

\subsubsection{Adaptive Combination of Node Features}

The node features learned using the graph network have to be fused in some form for the final classification to take place. Common choices from the literature include concatenation, addition, or a dot product. In this work, before combining the features finally through such a mechanism, we pass the features through an adaptive weight prediction module, similar to the attention mechanism~\cite{vaswani2017attention}. We denote this \underline{a}daptive \underline{c}ombination method as AC. The module works in the form of a single-layered feed-forward network, that predicts a single weight coefficient to be applied in a complementary manner to the node features of both the source and destination nodes of any edge being considered for classification. Mathematically, the weight \(\beta\) is calculated in the following way: 
\begin{equation}
    \beta = \exp(\text{LeakyReLU}(\mathbf{w}_{u}x_{u} + \mathbf{w}_{i}x_{i}))\;,
\end{equation}
where \(x_{u} \in \m{V}_{user}\) and \(x_{i} \in \m{V}_{image}\), and \(\mathbf{w}_{u}, \mathbf{w}_{i}\) are learnable scalar weights for the user and image node features respectively. We do not include the edge attributes to be adaptively combined as we initialize missing edge attributes uniformly. The randomly initialized feature tensors might not hold information reliable enough for the final classification step. 
Once the weight coefficient is obtained, the combination takes place as follows, 
\begin{equation}
    x_\text{comb} = c(\beta x_{u}, (1-\beta)x_{i})\;.
\end{equation}
Here, \(c(\cdot)\) denotes the final combination mechanism. 

\section{Experiments and Results}
\label{sec4}
\begin{table}
\caption{Statistics of the Created Graph}
\begin{center}
\renewcommand\arraystretch{1.2}
\begin{tabular}{|c|c|c|c|}
\hline
User Nodes & Image Nodes & User-User Edges & User-Image Edges \\
\hline\hline
108899 & 85157 & 1649058 & 197561 \\
\hline
\end{tabular}
\label{tab1}
\end{center}
\end{table}

\subsection{Dataset Description}
We use the Image-Emotion-Social-Net (IESN) dataset~\cite{zhao2016predicting} to validate the effectiveness of HMG-Emo. As the sole publicly available dataset for personalized emotion prediction in social networks, IESN stands out by including \textit{contexts about users} from the social media platform Flickr\footnote{https://www.flickr.com}, alongside images and ground-truth emotion labels for emotion prediction. The dataset includes eight emotion classes, based on Mikel's model~\cite{mikels2005emotional} - \textit{Amusement}, \textit{Anger}, \textit{Awe}, \textit{Contentment}, \textit{Disgust}, \textit{Excitement}, \textit{Fear}, and \textit{Sadness}, as well as an \textit{Unknown} emotion category. We use the following diverse information from the dataset: 
\begin{itemize}[left=0pt]
    \item \textit{Actual Emotion of User-Image Interaction}: This part of the dataset provides details on users, images, and their interaction such as the timestamp, along with ground truth emotion labels representative of the \textit{actual} emotion evoked in users by viewing the images. These ground truth labels are primarily derived from the user comments by calculating their average Valence, Arousal, and Dominance (VAD) using  VAD norms ~\cite{warriner2013norms}.
    \item \textit{User Group Information}: The dataset provides information about special interest groups on Flickr, where multiple users join in to share images related to a specific topic. This information includes descriptions of the group interests and a membership list. 
    \item \textit{User Information}: Additional user data such as the user contact lists are also available in this dataset. 
\end{itemize}
As IESN~\cite{zhao2016predicting} was introduced in 2016, some images and comments included in the dataset may no longer exist, or their parent user profiles may have been deleted. We remove such entries. A number of the user-image interactions also contain multiple related emotion labels or an emotion label of \textit{Unknown}. We use only a single ground truth emotion label for a user-image edge and remove instances that are labeled \textit{Unknown}. The comments are also presented only in the form of comment IDs, and the actual text is obtained directly from Flickr, using those IDs. Then, we construct the heterogeneous graph based on the user, image, contact list, and comment information. 
In case the actual comment text is missing, 
we initialize the corresponding edge attributes to be a uniform tensor. 
Table \ref{tab1} provides an overview of the statistics of the created graph. 
Among the 197,561 user-image edges, a large number of edges lack the original comment texts, leading to only 34\% comment features available extracted using mBERT~\cite{devlin2019bert}. 
Note that the number of user nodes depicted in Table \ref{tab1} can refer to users as either image uploaders, image viewers, or both. Still, they do not necessarily have to be both simultaneously. 

\begin{table}[t]
\caption{Emotion Classification Results}
\begin{center}
\resizebox{\columnwidth}{!}{
\renewcommand\arraystretch{1.2}
\begin{tabular}{|c|c|c|c|}
\hline
Method & F1 & Precision & Recall\\
\hline
\hline
RMTHG (V)~\cite{zhao2016predicting} & \(0.44 \pm 0.07\) & \(0.36 \pm 0.16\) & \(0.68 \pm 0.12\)\\
RMTHG (Fusion)~\cite{zhao2016predicting} & \(0.60 \pm 0.1\) & \(0.50 \pm 0.1\) & \(\mathbf{0.72 \pm 0.1}\) \\
HMG-Emo\(-\)\{T, AC\} & \(0.75 \pm 0.004\) & \(0.92 \pm 0.009\) & \(0.64 \pm 0.001\) \\
HMG-Emo\(-\)\{AC\} & \(0.76 \pm 0.003\) & \(0.94 \pm 0.005\) & \(0.64 \pm 0.002\) \\
\textbf{HMG-Emo} & \(\mathbf{0.77 \pm 0.003}\) & \(\mathbf{0.97 \pm 0.007}\) & \(0.65 \pm 0.001\) \\
\hline
\end{tabular}
}
\label{tab2}
\end{center}
\end{table}

\begin{table}[tbp]
\caption{Emotion Classification with Comments Generated by LLaVA}
\begin{center}
\resizebox{\columnwidth}{!}{
\renewcommand\arraystretch{1.2}
\begin{tabular}{|c|c|c|c|}
\hline
Method & F1 & Precision & Recall\\
\hline
\hline
HMG-\(\m{T}\) comments only & \(0.30 \pm 0.003\) & \(0.31 \pm 0.006\) & \(0.34 \pm 0.002\) \\
LLaVA comments only & \(0.36 \pm 0.002\) & \(0.38 \pm 0.003\) & \(0.38 \pm 0.002\) \\
\hline 
\end{tabular}
}
\label{tab3}
\end{center}
\end{table}

\subsection{Experimental Setup}

The implementations in all experiments are based on PyTorch Geometric~\cite{fey2019fast}. 
The models are trained on an NVIDIA A40 GPU node with four GPU cores. 
Given the dataset's significant class imbalance, we report the weighted F1 score as the classification performance metric. We make an 80:10:10 train-validation-test split and report 3-fold cross-validated scores on the test data, along with the standard deviation across total runs. We use an output dimension size of 64 for the graph network in HMG-Emo. The models are trained for 20 epochs using the Adam optimizer~\cite{kingma2014adam}, with a base learning rate of 0.005, and cosine annealing scheduler~\cite{loshchilov2016sgdr}. The batch size used for the experiments is 512. The embedding dimensions chosen for the initial features are 128 for the user, 128 for the image, and 256 for the comment features, chosen empirically. We use five layers of each graph neural network backbone, with the default choice for HMG-Emo being Graph Attention layers~\cite{velivckovic2018graph}. Convergence in training takes approximately 4 hours at the longest.

\subsection{Emotion Prediction Experiments}

\subsubsection{With IESN and HMG-Emo}

The primary results using HMG-Emo are presented in Table \ref{tab2}. We compare our proposed method with the only baseline in this problem, introduced in~\cite{zhao2016predicting}, which adopted a Rolling Multi-task Hypergraph Learning framework (RMTHG). 
RMTHG considered multiple factors, such as image features, similarity between users in terms of the comments they make, the groups they join, as well as interaction records such as the timestamps of comments. 
RMTHG predicts the emotional state of the user when they view a particular image, based on all of these contextual pieces of information. It however considers image features and user information to be part of a single complex vertex in the hypergraph. The variant of RMTHG where only visual features are used is denoted using RMTHG (V). Similarly, for HMG-Emo, the variant \textit{without} comment features or adaptive combination is denoted as HMG-Emo\(-\)\{T, AC\}, and the variant that uses all features - visual, user context, and comments, but not the adaptive combination, is denoted as HMG-Emo\(-\)\{AC\}. Our formulation of the problem achieves a significant boost in the performance of personalized emotion classification, with our complete framework achieving an improvement of 28\% over the multimodal baseline, and 75\% over the vision-only baseline. The use of multiple features within the graph network, along with an adaptive combination in the classification step further strengthens the model. HMG-Emo tends to have a very low rate of predicting false positives, leading to high precision, as opposed to the significantly high value of recall achieved by the RMTHG method. We speculate that the difference in precision and recall is due to the highly imbalanced nature of the data. The higher values of precision compared to recall signal better performance in the majority class, which can also be understood by the significantly lower number of samples present for the minority class. However, the precision and recall scores achieved by HMG-Emo can be considered substantially high enough for the model to be useful for downstream tasks. 

\subsubsection{Effect of Multimodal Features}

As described early in Section \ref{sec4}, the ground truth emotion labels for actual emotions of users, in IESN~\cite{zhao2016predicting}, are obtained primarily from the comments left by the users. This may spark a question as to whether the comments can only be considered sufficient for predicting personalized emotions in users. To verify, we have also conducted an experiment using only the comment features from HMG-Emo. For a fair comparison in classification performance, we use the same feedforward neural network architecture as in the HMG-Emo classification module, described in Section \ref{sec3-c}. The first row in Table \ref{tab3} shows the results, where HMG-\(\m{T}\) denotes the comment features originally used in HMG-Emo as edge attributes. The classification performance is far below what is obtained using a graph-based approach, underlining the need for a multimodal approach. 

\begin{figure}[t]
    \centering
    \includegraphics[scale=0.5]{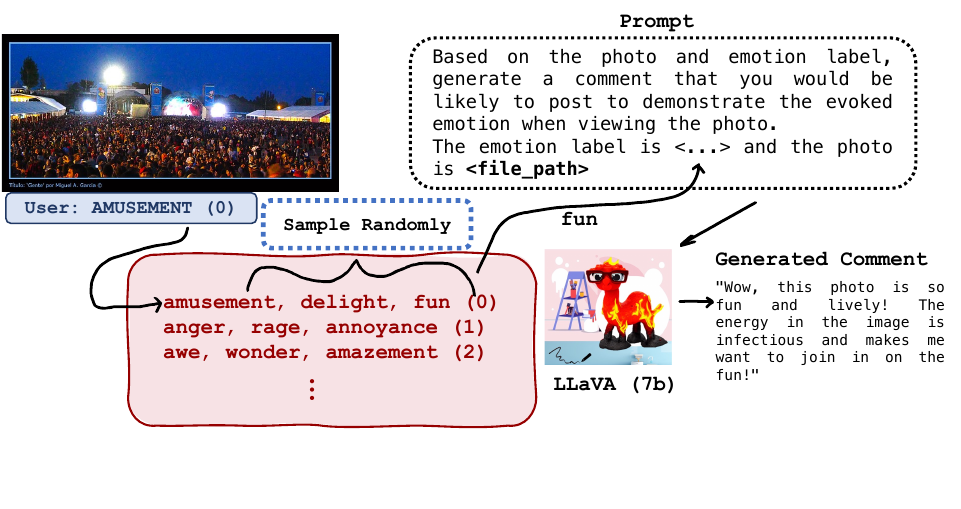}
    \caption{The prompting method used for LLaVA}
    \label{fig:4}
\end{figure}

However, we also consider that the original comment feature set HMG-\(\m{T}\) has only 34\% of the feature representations learned from actual comment texts, as the actual texts for others were missing, leading to them being filled with randomly sampled feature tensors. To delve deeper and verify whether that is the reason for the poor classification performance, we employ LLaVA (7b)~\cite{liu2023visual} to generate actual texts for the missing comments. LLaVA is chosen because it is open-source, and relatively lightweight, making it easier to reproduce the results with minimal resource constraints. For every edge that is missing the original comment features, we use the corresponding images, and ground truth emotion labels to prompt LLaVA to generate comments that are representative of the ground truth emotion. The prompt thus depends solely on the image, and the associated evoked emotion label. This presents a challenge, as multiple edges exist between the same image, and different users, often having the same emotion label. This could lead to the same prompt being used for multiple users, leading to identical comments being generated for different users viewing the same image. To ensure that the comments generated by LLaVA for different users are unique, randomly sampled synonyms of the emotion label name are used in the prompts. The entire process is depicted in Fig. \ref{fig:4}. We manually and qualitatively validate a subset of such generations from LLaVA and find them to be highly correlated with the corresponding emotion labels. The comments often contain keywords directly related to the ground truth emotion label, which makes them significantly richer than real-world comments, which may often contain unrelated or noisy information. The same feature extraction method of mBERT~\cite{devlin2019bert} is used on the comments generated using LLaVA, and emotion classification is carried out. As can be noted from Table \ref{tab3}, even with rich features learned from comments that are highly related to the actual emotion label, the classification performance is largely below par. This stresses the need for using a framework that learns from multiple modalities in tandem, to achieve a holistic understanding of user emotions in social networks. 

\subsubsection{Graph Backbone Modification}

We experiment with different backbone graph layers to understand the impact of using a specific graph layer in our model. We compare our model with the counterparts obtained when the Graph Attention layers in them are replaced by GraphSAGE~\cite{hamilton2017inductive} and GraphConv~\cite{morris2019weisfeiler} layers. Both GraphSAGE and GraphConv do not however take into consideration the edge attributes by default. Thus, for a fair comparison, we compare them with HMG-Emo with the comment features removed. Note that this still includes the adaptive combination of user and image modalities, hence it is different from HMG \(-\)\{T, AC\}. From Table \ref{tab4}, we can observe that with our framework, there is not much variation in the emotion prediction performance, even when the backbone layers are changed, demonstrating that the improvement in performance does not only come from the choice of graph attention as the backbone. 



\begin{table}[tbp]
\caption{Ablation Study for Different Graph Backbone}
\begin{center}
\renewcommand\arraystretch{1.2}
\begin{tabular}{|c|c|c|c|}
\hline
Graph Backbone & F1 & Precision & Recall\\
\hline
\hline
GAT  & \(0.77 \pm 0.002\) & \(0.98 \pm 0.004\) & \(0.65 \pm 0.0002\) \\
GraphSAGE & \(0.75 \pm 0.004\) & \(0.92 \pm 0.012\) & \(0.65 \pm 0.001\) \\
GraphConv & \(0.76 \pm 0.001\) & \(0.93 \pm 0.006\) & \(0.66 \pm 0.002\) \\
\hline 
\end{tabular}
\label{tab4}
\end{center}
\end{table}

\begin{table}[tbp]
\caption{Ablation Study for Negative Sampling in Training and Inference}
\begin{center}
\resizebox{\columnwidth}{!}{
\renewcommand\arraystretch{1.2}
\begin{tabular}{|c|c|c|c|}
\hline
Method & F1 & Precision & Recall\\
\hline
\hline
W/ Negative Sampling & \(0.71 \pm 0.002\) & \(0.80 \pm 0.002\) & \(0.64 \pm 0.002\) \\
W/o Negative Sampling & \(0.75 \pm 0.004\) & \(0.92 \pm 0.009\) & \(0.64 \pm 0.001\) \\
\hline 
\end{tabular}
}
\label{tab5}
\end{center}
\end{table}

\subsubsection{Negative Sampling Strategies}

Negative sampling within a graph involves generating potential edges that are not present in the original graph.
In our framework, we study the impact of creating negative edges. With negative sampling, the model is trained to classify edges into the 8 emotion classes while also learning to discern some edges as non-existent. 
This strategy leads to a slight drop in the performance of the model due to the ambiguity of generating negative samples in a dynamic social network. In this case, some non-existent edges can potentially become true edges in the future, making the decision on the existence of edges more challenging. 
Table \ref{tab5} compares the performance of our simplest model (HMG\(-\)\{T, AC\}) with and without negative sampling.


\section{Conclusion}
\label{sec5}

We present a graph and deep learning-based framework for personalized emotion prediction in social networks. We demonstrate significant gains in performance when utilizing deep features and multiple modalities simultaneously. {In conclusion, several promising directions can be considered for future research. 
Firstly, the proposed framework in this paper can be further enhanced by incorporating additional contexts, such as geo-tags of images and timestamps of posting images or comments, alongside adopting more sophisticated feature extraction techniques. With our method being easily extensible to new modalities of information, the inclusion of diverse data can be seamless. 
Moreover, given the increasing popularity of Large Language Models (LLMs) and Graph-LLMs, there exists an opportunity to input data into Graph-LLMs for emotion prediction directly.
Addressing the challenge of integrating multiple contexts into prompts for such LLMs remains an open area of inquiry. 
We hope that the insights from this study will inspire future research endeavors in this field.}

\section{Ethical Impact Statement}\label{sec6}
Our research introduces a novel framework for personalized emotion prediction in social networks, leveraging graph and deep learning techniques. While our work contributes to the advancement of affective computing, we recognize the importance of addressing ethical considerations in this domain. 

The proposed automatic pipeline and model raise concerns about algorithm transparency and accountability. We acknowledge the challenge of reproducing the same results in the experiments due to the complexity of deep learning models and the reliance on different hyperparameters. 
Therefore, we provide comprehensive details of the proposed method in this research and plan to make our codes for data collection and experiments publicly available upon acceptance.
Additionally, we emphasize that it is necessary to carry out strict evaluation to ensure the reliability and fairness of our method. This includes conducting experiments over multiple runs and providing corresponding error estimates. 

Furthermore, the utilization of social media data brings concerns regarding data privacy and user consent. We understand that it is pivotal to address user privacy rights and obtain explicit consent for the collection and analysis of sensitive user data. We confirm that we only use content that is already publicly available from social media and do not infringe upon the privacy of users on the social media platform. 

In a supplementary experiment utilizing a Large Language Model (LLM) to generate missing data, we acknowledge that the data includes the biases of the training set for the particular model. 
We also note that data generated from the LLM for our purpose is not representative of the real-world data quality. Based on this observation, we refrain from training our framework on the augmented dataset to avoid potential misleading performance gains obtained from idealistic data.

Moreover, we note that although this research aims to predict user emotions in social networks for social good, we should not rule out the possibility of the research being used for a malicious purpose, such as manipulating content on social media based on user emotions. Therefore, we advocate for the responsible usage of our research and similar endeavors, emphasizing thorough risk analysis it can pose to users on social media and its usage only in well-charted scenarios that prioritize user welfare.

\section*{Acknowledgments}

The work was supported in part by the National Science Foundation (NSF) under Grant No. 2234195. This work used cluster computers at the National Center for Supercomputing Applications through an allocation from the Advanced Cyberinfrastructure Coordination Ecosystem: Services \& Support (ACCESS) program, which is supported by NSF Grants Nos. 2138259, 2138286, 2138307, 2137603, and 2138296. We would also like to express our appreciation to the anonymous reviewers for their constructive feedback.

Some images are incorporated in figures for illustrative purposes and to support the conceptual discussions in this paper. All copyrights remain with their respective owners. The authors acknowledge and appreciate the work of the creators.

\bibliography{ref.bib}
\bibliographystyle{ieeetr}

\end{document}